\documentclass[showpacs,twocolumn,aps,floatfix]{revtex4-1}%
\usepackage{amssymb}
\usepackage{amsfonts}
\usepackage{amsmath}
\usepackage{graphicx}%
\providecommand{\U}[1]{\protect\rule{.1in}{.1in}}
\providecommand{\U}[1]{\protect\rule{.1in}{.1in}}

\begin{document}
\preprint{cond-mat}
\title[Short title for running header]{Non differentiable large-deviation functionals in boundary-driven diffusive systems}
\author{Guy Bunin}
\affiliation{Technion -- Israel Institute of Technology, Haifa 32000, Israel}
\author{Yariv Kafri}
\affiliation{Technion -- Israel Institute of Technology, Haifa 32000, Israel}
\author{Daniel Podolsky}
\affiliation{Technion -- Israel Institute of Technology, Haifa 32000, Israel}

\begin{abstract}
We study the probability of arbitrary density profiles in conserving
diffusive\ fields which are driven by the boundaries. We demonstrate the
existence of singularities in the large-deviation functional, the direct analog
of the free-energy in non-equilibrium systems. These singularities are unique
to non-equilibrium systems and are a direct consequence of the breaking of
time-reversal symmetry. This is demonstrated in an exactly-solvable model and
also in numerical simulations on a boundary-driven Ising model. We argue that
this singular behavior is expected to occur in models where the
compressibility has a deep enough minimum. The mechanism is explained using a
simple model.

\end{abstract}
\volumeyear{year}
\volumenumber{number}
\issuenumber{number}
\eid{identifier}
\startpage{1}
\endpage{4}
\maketitle

Consider a field $\rho(x)$ with diffusive dynamics which are conserving in the
bulk. Here, $\rho(x)$ could describe the density of a gas in a capillary
connecting two reservoirs. When the reservoirs are of equal density $\bar
{\rho}$, the system is in equilibrium. Then the steady-state probability
$P[\rho_{f}]$ of an arbitrary density profile $\rho_{f}(x)$ is given by
$P[\rho_{f}]\sim e^{-F[\rho_{f}]/k_{B}T}$, where $F$ is the free-energy.
Generically, for a system with short-range interactions, $F$ is a local
functional of $\rho_{f}(x)$ and, in the disordered phase, it is a smooth
functional. For instance, when the particles in the capillary interact only
through hard-core exclusion, $F[\rho_{f}]=N\int_{0}^{1}dx\,\{\rho_{f}%
(x)\log\frac{\rho_{f}(x)}{\bar{\rho}}+(1-\rho_{f}(x))\log\frac{1-\rho_{f}%
(x)}{1-\bar{\rho}}\}$, where $N$ is the length of the capillary and the
density is normalized such that $\rho=1$ corresponds to a filled capillary.

A fundamental goal of non-equilibrium statistical mechanics is to evaluate and
understand the general structure of $P[\rho_{f}]$ when the densities of the
two reservoirs are different, such that a current flows through the system.
For diffusive systems there has been great progress in recent years, and by
now several important properties of $P[\rho_{f}]$ have been established. For
example, it is known that $P[\rho_{f}]\sim e^{-N^{d}\phi\lbrack\rho_{f}]}$,
where $N^{d}$ is the volume of the system and $\phi\lbrack\rho_{f}]$ is a
non-equilibrium analogue of the free energy, called the large deviation
functional (LDF). It attains a minimum at the most probable density profile
and, when the system is driven out of equilibrium, it becomes a
\emph{non-local functional} of $\rho_{f}\left(  x\right)  $. This has
important consequences, manifested for example through long range correlations
which are present even when the interactions are strictly local
\cite{SSEP_spohn,Derrida_review}. In contrast to equilibrium the LDF depends
on the dynamics, and not only on the Boltzmann weights.

Recently, framework within which $\phi\lbrack\rho_{f}]$ can be calculated has
been laid out, building on standard tools from the theory of large deviations
\cite{BertiniPRL,BertiniJstat,JSP_quantum,TKL_long,Freidlin_Wentzell,Touchette}%
. The key observation is that in these systems the probability of an atypical
event is dominated by a single history leading up to it, and starting from the
most probable profile; other histories are exponentially less likely in the
system size. While calculating the LDF within the framework is in general
extremely difficult, it has allowed to establish exact solutions in a number
of simple models \cite{BertiniJstat,KMP_large_dev} (in some cases building on
solutions obtained by other methods \cite{SSEP_large_dev}). In addition, the
framework has led to efficient numerical algorithms for evaluating $P[\rho
_{f}]$ for arbitrary diffusive models \cite{our_numerics}, as well as LDFs of
global quantities such as the current
\cite{Giardina,TV_simulation_cont_t,current_large_dev_review,Giardina_review}.

Despite the successes, a general understanding of the properties of
$P[\rho_{f}]$ out of equilibrium is lacking. For example, for boundary-driven
currents induced by reservoirs held at different densities, $\phi\lbrack
\rho_{f}]$ is a smooth functional of $\rho_{f}$ for particles diffusing with
hard-core exclusion, just as in equilibrium \cite{SSEP_large_dev}. In stark
contrast, in the presence of a strong bulk drive (for example, when a constant
force acts on the particles in the capillary)\textbf{ }$\phi\lbrack\rho_{f}]$
becomes non-analytic \cite{ASEP_transition}. This is due to the existence of
multiple histories leading to the same $\rho_{f}$ with comparable weight, and
immediately implies singular behavior in the LDF of other more global
quantities.\ These singularities are more subtle than those observed for the
current or particle number
\cite{Bertini_current_phase_trans,Bodineau,Merhav_kafri,hurtado_current_KMP},
which can arise even if the LDF on the phase space is smooth, hence even in
equilibrium models (in contrast to the present phenomenon, see below). It is
far from clear which models exhibit such singular behavior, and how to
characterize these singularities. It is natural to ask whether these
singularities can appear when no bulk drive is present.

In this Letter we study boundary-driven diffusive systems and show for the
first time that singular behavior in $\phi$ can indeed occur in these systems.
We first provide an example of an exactly solvable model which exhibits such a
singularity. Then we use numerics to demonstrate that this phenomenon occurs
in a broad class of models (for example, in a boundary-driven Ising model). We
elucidate the mechanism by which the singularities arise, by introducing a
simple model which contains the essential ingredients leading to the
singularity. This model shows the close connection between the singularities
and a spontaneous symmetry breaking. We provide guidelines to systems which
can be expected to show this singular behavior. Throughout our analysis we use
analogies with previous works on the Fokker-Planck equation in the presence of
small noise. Singular behaviors in such models have been studied extensively
in the past
\cite{Graham_Tel,Noise_non_lin_book,analog_exp,Maier_Stein,Dykman,Dykman_cusp_structure}%
, and we draw analogies with these works whenever possible.

\emph{Background theory} -- We study large deviations of bulk-conserving
diffusive systems which are driven out of equilibrium by the boundaries. For
simplicity we consider one dimension, with $0\leq x\leq1$. The conserved
density $\rho\left(  x,t\right)  $ is related to the current $J\left(
x,t\right)  $ by $\partial_{t}\rho+\partial_{x}J=0$, where the current is
given by
\begin{equation}
J=\mathbf{-}D\left(  \rho\left(  x,t\right)  \right)  \partial_{x}\rho\left(
x,t\right)  +\sqrt{\sigma\left(  \rho\left(  x,t\right)  \right)  }\eta\left(
x,t\right)  \ .\label{eq:J_def}%
\end{equation}
$D\left(  \rho\left(  x,t\right)  \right)  $ is a density-dependent
diffusivity function, while $\sigma\left(  \rho\left(  x,t\right)  \right)  $
controls the amplitude of the white noise $\eta\left(  x,t\right)  $, which
satisfies $\left\langle \eta\left(  x,t\right)  \right\rangle =0$ and
$\left\langle \eta\left(  x,t\right)  \eta\left(  x^{\prime},t^{\prime
}\right)  \right\rangle =N^{-1}\delta\left(  x-x^{\prime}\right)
\delta\left(  t-t^{\prime}\right)  $. The prefactor $N^{-1}$ in the noise
variance results from the fact that we have scaled distances by the system
size $N$, and time by $N^{2}$. After this rescaling the noise is small due to
the system size, as a consequence of the coarse graining and rescaling. Eq.
(\ref{eq:J_def}) describes a broad range of transport phenomena, including
electronic systems, ionic conductors, and heat conduction
\cite{JSP_quantum,ionic_conductors,KMP}. $D\left(  \rho\right)  $ and
$\sigma\left(  \rho\right)  $ are related via a fluctuation-dissipation
relation, which for particle systems reads $\sigma\left(  \rho\right)
=2k_{B}T\rho^{2}\kappa\left(  \rho\right)  D\left(  \rho\right)  $ where
$\kappa\left(  \rho\right)  $ is the compressibility \cite{Derrida_review}.
For example, for diffusing hard-core particles, $D=1$ and $\sigma=2\rho\left(
1-\rho\right)  $ \cite{Derrida_review}. The system is attached to reservoirs
which fix the densities at\ the boundaries of the segment, $\rho\left(
0\right)  =\rho_{L}$ and $\rho\left(  1\right)  =\rho_{R}$. If $\rho_{L}%
\neq\rho_{R}$ a current is induced through the system, driving it out of equilibrium.

The average, or most probable density profile for the system $\bar{\rho}$, is
obtained by solving $\partial_{x}\left[  D\left(  \bar{\rho}\right)
\partial_{x}\bar{\rho}\right]  =0$, with $\bar{\rho}\left(  0\right)
=\rho_{L}$ and $\bar{\rho}\left(  1\right)  =\rho_{R}$ at the boundaries. In
equilibrium (i.e. when $\rho_{L}$ $=\rho_{R}$), the steady-state probability
of any other density profile $\rho\left(  x\right)  $ is easy to obtain -- the
LDF $\phi\left[  \rho\right]  $ is then given by the free-energy which is
local in $\rho$, $\phi\lbrack\rho]=\int f\left(  \rho,\bar{\rho}\right)  dx$,
where
\begin{equation}
f\left(  \rho,r\right)  \equiv\int_{r}^{\rho}d\rho_{1}\int_{r}^{\rho_{1}}%
d\rho_{2}\frac{2D\left(  \rho_{2}\right)  }{\sigma\left(  \rho_{2}\right)
}\ . \label{eq:free_energy_density}%
\end{equation}
Note that in this case $\bar{\rho}$ is constant, $\bar{\rho}=\rho_{L}$
$=\rho_{R}$. By contrast, the steady-state probability distribution away from
equilibrium is notoriously hard to compute, and \emph{very} different from the
naive guess $\phi\lbrack\rho]=\int f\left(  \rho,\bar{\rho}\right)  dx$, now
with space dependent $\bar{\rho}\left(  x\right)  $. In fact, as stated above,
$\phi\lbrack\rho]$ is non-local.

To compute the large deviation for the model described above, one uses the
fact that the probability of a history $\left\{  \rho\left(  x,t\right)
,J\left(  x,t\right)  \right\}  $\ during time $-\infty\leq t\leq0$ is
$P\sim\exp\left(  -NS\right)  $, where the action $S$ is given by
\cite{BertiniPRL,BertiniJstat,JSP_quantum,TKL_long,Freidlin_Wentzell}
\begin{equation}
S=\int_{-\infty}^{0}dt\int_{0}^{1}dx\frac{\left[  J\left(  x,t\right)
+D\left(  \rho\left(  x,t\right)  \right)  \partial_{x}\rho\left(  x,t\right)
\right]  ^{2}}{2\sigma\left(  \rho\left(  x,t\right)  \right)  }\ .
\label{eq:action}%
\end{equation}
For large $N$, the probability $P\sim e^{-N\phi\left[  \rho_{f}\right]  }$ is
given by $\phi\left[  \rho_{f}\right]  =\inf_{\rho,J}S$, where the infimum is
over histories satisfying $\partial_{t}\rho+\partial_{x}J=0$, with initial and
final conditions $\rho\left(  x,t\rightarrow-\infty\right)  =\bar{\rho}\left(
x\right)  $, $\rho\left(  x,t=0\right)  =\rho_{f}\left(  x\right)  $, and
boundary conditions $\rho\left(  x=0,t\right)  =\rho_{L}$ and $\rho\left(
x=1,t\right)  =\rho_{R}$.

In many cases, including in equilibrium and in previously studied exactly
solvable non-equilibrium models \cite{SSEP_large_dev,KMP_large_dev}, the
action $S$\ in Eq.~(\ref{eq:action})\ has a single local minimum and
$\phi\left[  \rho_{f}\right]  $ is then a smooth functional. However, as
we\ show below this need not be the case.\textbf{ }The action can, in general,
have more than one local minimum in the space of histories $\left\{
\rho\left(  x,t\right)  ,J\left(  x,t\right)  \right\}  $. \textbf{\ }In
regions of the space of final states where $\rho_{f}$ has more than one
minimal history leading to it, the global minimum might switch between two
locally minimizing solutions. This is analogous to a first-order phase
transition in equilibrium statistical mechanics, where the system switches
between two metastable states, which are both local minima of the free-energy.
The transition between local minima is accompanied by a jump in the functional
derivative of the large-deviation $\delta\phi/\delta\rho$. We will therefore
refer to it as a Large Deviation Singularity (LDS). This phenomenon, first
studied by Graham and T\'{e}l \cite{Graham_Tel}, is unique to non-equilibrium:
in equilibrium $\phi$ is smooth whenever the dynamical model (Langevin
equation) contains only smooth functions. Note that while LDSs are expected to
be generic in models where the zero-noise dynamics feature a number of basins,
or unstable fixed points, here the only fixed point is $\bar{\rho}\left(
x\right) $, and therefore the existence of the singularity is not guaranteed,
and indeed is not present in the previously studied models
\cite{SSEP_large_dev,BertiniJstat,KMP_large_dev}.

We first study this phenomenon in an exactly solvable model, and then consider
its generalizations.

\emph{Exactly solvable model }-- Consider the model in Eq.~(\ref{eq:J_def})
with $D=1$ and a quadratic $\sigma\left(  \rho\right)  =1+\rho^{2}$, a
parabola clear above the axis \cite{all_quad_sig}. In \cite{KMP_large_dev} it
was shown that the LDF is given by $\phi\left[  \rho_{f}\right]  =\min
\phi_{ext}$, where $\phi_{ext}$ are extremal values of the action given by%
\[
\phi_{ext}=\int_{0}^{1}dx\left\{  f\left(  \rho_{f}\left(  x\right)  ,g\left(
x\right)  \right)  -\ln\frac{g^{\prime}\left(  x\right)  }{\bar{\rho}^{\prime
}\left(  x\right)  }\right\}
\]
where $f\left(  \rho,g\right)  $ is defined in Eq.
(\ref{eq:free_energy_density}) and $g\left(  x\right)  $ is an auxiliary
function satisfying the differential equation%
\begin{equation}
0=\frac{g\left(  x\right)  -\rho_{f}\left(  x\right)  }{\sigma\left(  g\left(
x\right)  \right)  }-\frac{g^{\prime\prime}\left(  x\right)  }{\left[
g^{\prime}\left(  x\right)  \right]  ^{2}}\ , \label{eq:general_ODE}%
\end{equation}
with boundary conditions $g\left(  0\right)  =\rho_{L}$, and $g\left(
1\right)  =\rho_{R}$.\ Note that as $D=1$, the most probable configuration
$\bar{\rho}\left(  x\right)  $ is linear, with $\bar{\rho}\left(  0\right)
=\rho_{L}$ and $\bar{\rho}\left(  1\right)  =\rho_{R}$. As we now show,
solutions to the differential equation Eq.~(\ref{eq:general_ODE}) with initial
and final boundary conditions may be non-unique, i.e. there exist profiles
$\rho_{f}\left(  x\right)  $ for which more than one solution $g\left(
x\right)  $ exists. Eq.~(\ref{eq:general_ODE}) is solved via a numerical
shooting method \cite{numerical_recepies}: here we treat the problem as an
initial value problem with initial conditions $g\left(  0\right)  =\rho_{L}$,
and $g^{\prime}\left(  0\right)  =c$, and scan systematically over values of
$c$ to find solutions with $g\left(  1\right)  =\rho_{R}$. This type of
exhaustive search ensures that all extremal states are discovered.

\begin{figure}
[ptb]
\begin{center}
\includegraphics[
height=1.6587in,
width=2.982in
]%
{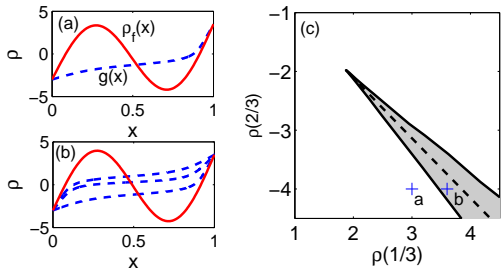}%
\caption{The model $D=1,\sigma=\rho^{2}+1$. (a) A profile $\rho_{f}\left(
x\right)  $ (solid line) with one extremal solution $\phi\left[  \rho\right]
$. The corresponding $g\left(  x\right)  $ also plotted\ (dashed line). (b) A
profile $\rho_{f}\left(  x\right)  $ with three extremal solutions. (c) Number
of extremal solutions as a function of $\rho_{f}\left(  1/3\right)  ,\rho
_{f}\left(  2/3\right)  $. Gray region: three solutions. White region: one
solution. Also shown are the caustics (solid line), and the switching line
(dashed line).}%
\label{fig:exact_model}%
\end{center}
\end{figure}

To illustrate the existence of multiple solutions, we consider profiles of the
form $\rho_{f}\left(  x\right)  =\bar{\rho}\left(  x\right)  +\alpha_{1}%
\cos\left(  \pi x/2\right)  +\alpha_{2}\sin\left(  \pi x\right)  $, varying
$\alpha_{1}$ and $\alpha_{2}$, and with boundary-conditions $\rho_{L}%
=-3,\rho_{R}=3$. Fig. \ref{fig:exact_model}(a) shows an example of a profile
$\rho_{f}$ for which only one solution $g\left(  x\right)  $ to Eq.
(\ref{eq:general_ODE}) exists. In contrast, in Fig. \ref{fig:exact_model}(b) a
profile $\rho_{f}\left(  x\right)  $ with three solutions $g\left(  x\right)
$ is shown, two of them corresponding to local minima, and one to a saddle
point. Fig. \ref{fig:exact_model}(c) shows the region in which there are
multiple solutions. Here we have chosen to parametrize the profiles $\rho
_{f}\left(  x\right)  $ in terms of $\rho_{f}\left(  1/3\right)  $ and
$\rho_{f}\left(  2/3\right)  $, which are simply related to $\alpha_{1}$ and
$\alpha_{2}$. Note the marked \emph{caustics}, indicating the boundaries
between regions with one and three extremal solutions, and the \emph{switching
line}, on which the two locally minimal solutions have the same value for the
action.\ On the switching line the gradient of the LDF is discontinuous
\cite{Graham_Tel}; the occurrence of this LDS is the focus of the paper. In
addition, the history preceding a rare event is expected to be different on
both sides of the switching line, as the history minimizing the action
changes. The cusp is found for profiles $\rho_{f}$ relatively
\textquotedblleft far\textquotedblright\ from $\bar{\rho}$, and for $\rho
_{f}\left(  1/3\right)  $ positive and$\ \rho_{f}\left(  2/3\right)  $
negative (here $\rho_{L}<\rho_{R}$). More generally, the phase space of
profiles $\rho_{f}$ is infinite dimensional, the caustics and switching line
become manifolds. The picture shown in Fig. \ref{fig:exact_model}(c) is a
particular two dimensional cross section. We find similar behavior for
$\rho_{f}$ of similar shapes which are not of the exact form described above.

In fact, for this model one can prove that: (a) for \emph{any} non-equilibrium
boundary conditions ($\rho_{L}\neq\rho_{R}$) there exist profiles $\rho_{f}$
for which there is more than one solution to Eq.~(\ref{eq:general_ODE}), and
(b) profiles which have two locally minimizing histories with the same value
of $\phi_{ext}$ always exist.\ A proof of these facts will be given elsewhere
\cite{ours_long}.%

The existence of multiple minimizers of $S$ can be intuitively understood as
follows: looking at Eq.~(\ref{eq:action}), we see that the contribution to the
action is smaller wherever $\sigma\left(  \rho\right)  $ is high. If the
variation in $\sigma\left(  \rho\right)  $ is large enough, trajectories
passing through regions of high $\sigma\left(  \rho\right)  $ may be favored.
If there are two such regions, as around a minimum of $\sigma$, there might be
different paths of the action utilizing the different favorable $\sigma\left(
\rho\right)  $ regimes. When $D\left(  \rho\right)  $ also varies, we expect
the same logic to apply to regions with high and low $\sigma\left(
\rho\right)  /D\left(  \rho\right)  $. This argument suggests that the
phenomenon is robust, and will occur in other modes with similar features,
i.e. when $\sigma\left(  \rho\right)  /D\left(  \rho\right)  $ has a
pronounced minimum. Below we make this argument precise, but first we
demonstrate the generality of the phenomenon by studying it on a different
model which admits a concrete microscopic realization.

\begin{figure}
[ptb]
\begin{center}
\includegraphics[
height=2.1775in,
width=3.1355in
]%
{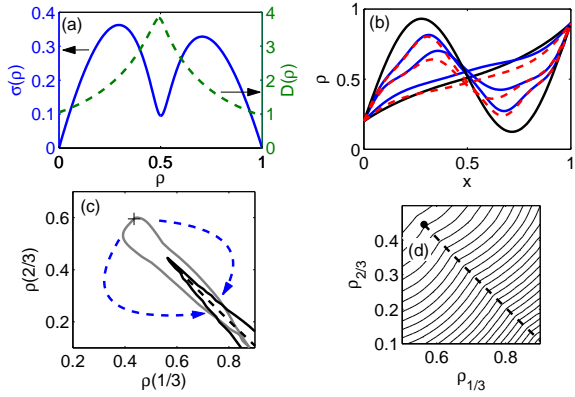}%
\caption{The boundary-driven Ising model. (a) $D\left(  \rho\right)  $ and
$\sigma\left(  \rho\right)  $. (b) Evolution of two locally-minimizing
histories leading to the same $\rho_{f}$. (c) The coexistence region, showing
the caustics (solid black line), switching line (dashed line), and
cross-sections of two histories (gray line). Dashed arrows depict paths of the
final state in the numerics which yield different local minimizers. (d)
Contours of equal LDF (solid lines) and the switching line (dashed).}%
\label{fig:KLS}%
\end{center}
\end{figure}

\emph{Boundary driven Ising model} -- We turn to study a boundary-driven Ising
model, a lattice gas with on-site exclusion and nearest-neighbor interaction.
(This is a variant of the\ Katz-Lebowitz-Spohn model \cite{KLS}, but with no
bulk bias.) Each site $i=1,..,N$ of a one-dimensional lattice can be either
occupied (\textquotedblleft1\textquotedblright) or empty (\textquotedblleft%
0\textquotedblright). The jump rate from site $i$ to site $i+1$ depends on the
occupation at sites $i-1$ to $i+2$ according to the following rules:
$0100\overset{1+\delta}{\rightarrow}0010,1101\overset{1-\delta}{\rightarrow
}1011,1100\overset{1+\varepsilon}{\rightarrow}1010,1010\overset{1-\varepsilon
}{\rightarrow}1100$, and their spatially inverted counterparts with identical
rates. The parameter $0<\varepsilon<1$ corresponds to attractive interactions
between the particles; $\delta$ controls the density dependence of the
mobility. As shown in \cite{spohn_book,KLS_sigma}, for each parameter set
$\left(  \varepsilon,\delta\right)  $ one can write implicit analytic
equations for $D\left(  \rho\right)  $ which can then be inverted numerically.
Then $\sigma\left(  \rho\right)  $ is obtained via the fluctuation-dissipation
relation. Fig. \ref{fig:KLS}(a) shows $D\left(  \rho\right)  $ and
$\sigma\left(  \rho\right)  $ for $\left(  \varepsilon,\delta\right)  =\left(
0.05,0.995\right)  $. For equilibrium boundary conditions this model admits an
Ising measure.

We now solve for local minimizers of the action $S$, using the numerical
technique described in \cite{our_numerics}. The numerical solutions are
obtained by gradually changing\ the end profile $\rho_{f}$, while continuously
maintaining a locally-minimizing history $\rho\left(  x,t\right)  $. Different
locally-minimizing solutions are obtained by changing the final profile to
enter the bi-stable region from different directions, see Fig. \ref{fig:KLS}%
(c). As for the exactly-solvable model, we again find configurations $\rho
_{f}\left(  x\right)  $ for which multiple local minimizers of $S$ exist. In
Fig. \ref{fig:KLS}(b), we show two different histories $\rho\left(
x,t\right)  $\ leading to the same profile $\rho_{f}$, which is chosen again
to be of the form: $\rho_{f}=\bar{\rho}\left(  x\right)  +\alpha_{1}\sin\pi
x+\alpha_{2}\sin2\pi x$. In Fig. \ref{fig:KLS}(c) we plot the
numerically-obtained region in the $\rho_{f}\left(  1/3\right)  ,\rho
_{f}\left(  2/3\right)  $ plane with multiple minimizers, together with the
\textquotedblleft switching line\textquotedblright\ and the caustics. In
addition, for a particular $\rho_{f}\left(  x\right)  $ we depict the two
histories leading to it by showing the time-dependent values of $\rho\left(
x=2/3,t\right)  $ against $\rho\left(  x=1/3,t\right)  $. Lines of equal LDF
are plotted in Fig. \ref{fig:KLS}(d) to show the jump in its gradient.%

Note that in this model and for the chosen parameters, $\sigma\left(
\rho\right)  /D\left(  \rho\right)  $ has a double-hump structure with a deep
minimum, so an LDS is expected from the simple considerations discussed above.
Indeed, we have experimented with various forms of $D\left(  \rho\right)  $
and $\sigma\left(  \rho\right)  $, and conclude that an LDS occurs when
$\sigma\left(  \rho\right)  /D\left(  \rho\right)  $ has a minimum which is
deep enough. Recall that by fluctuation-dissipation is related to the
compressibility $\sigma\left(  \rho\right)  /D\left(  \rho\right)
=2k_{B}T\rho^{2}\kappa\left(  \rho\right)  $. These features have to be large
enough for this to happen; small changes to a model which does not feature an
LDS are generally not enough. When LDSs do appear, we have always found them
in regions of phase space where the profiles $\rho_{f}\left(  x\right)  $ have
a similar shape to that shown in Fig. \ref{fig:exact_model}(b) and Fig.
\ref{fig:KLS}(b). The exact conditions and profiles for an LDS to appear must
be studied on a model-specific basis. Below we provide a simple model which
illustrates the mechanism leading to LDSs in models with these features.

\emph{Mechanism} -- Finally we elucidate the connection between a deep minimum
in $\sigma\left(  \rho\right)  $ and the LDS in the configurations discussed
above. To do so we introduce a simple sweater sleeve model. We consider a
model with $D=1$ and $\sigma\left(  \rho\right)  $ of order one everywhere
except for a narrow range of density values, where it is small:%
\[
\sigma\left(  \rho\right)  =\left\{
\begin{array}
[c]{ccc}%
\sigma_{0}\varepsilon^{2} &  & \rho\in\left[  -\varepsilon/2,\varepsilon
/2\right] \\
\sigma_{1}\left(  \rho\right)  &  & otherwise
\end{array}
\right.  \ .
\]
Here $\varepsilon$ is a small parameter, and $\sigma_{1}\left(  \rho\right)  $
is some function independent of $\varepsilon$. Then $\sigma\left(
\rho\right)  $ has a deep minimum around $\rho=0$. \ (The fact that
$\sigma\left(  \rho\right)  $ is discontinuous is not essential -- a smoothed
version can also be used.)%

\begin{figure}
[ptb]
\begin{center}
\includegraphics[
height=2.0241in,
width=2.6308in
]%
{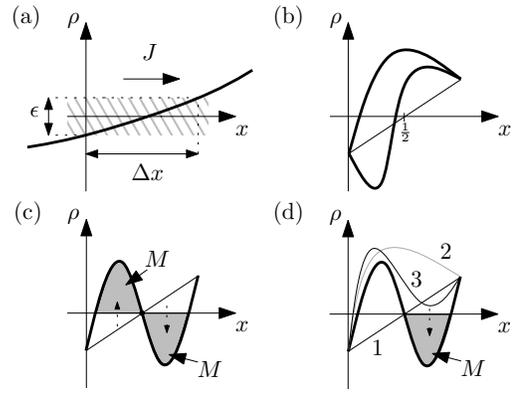}%
\caption{(a) Passing a mass through the noise barrier $\rho\in\left[
-\varepsilon/2,\varepsilon/2\right]  $. (b) Two profiles which can be reached
by only pushing mass \textquotedblleft downhill\textquotedblright\ through the
noise barrier. (c) A symmetric history leading to $\rho_{f}$ requires passing
a mass $2M$, shown in gray, \textquotedblleft uphill\textquotedblright. (d) A
symmetry-breaking history which requires passing only a mass $M$
\textquotedblleft uphill\textquotedblright. In (b-d) the straight thin line
depicts $\bar{\rho}$, and the bold lines $\rho_{f}$.}%
\label{fig:sweater_sleeve}%
\end{center}
\end{figure}
The key to estimating the action is noting that pushing a mass through the
density region $\rho\in\left[  -\varepsilon/2,\varepsilon/2\right]  $ may be
costly in terms of the action, creating a \textquotedblleft noise
barrier\textquotedblright. Consider the cost of passing a small mass element
$m$ from $\rho=-\varepsilon/2$ to$\ \rho=\varepsilon/2$ or vice versa, by
applying a current $J$, see Fig. \ref{fig:sweater_sleeve}(a). This process is
done over a time $\Delta t$. The region of space where $\rho\in\left[
-\varepsilon/2,\varepsilon/2\right]  $ is of length $\Delta x$. As $J=m/\Delta
t,\partial_{x}\rho=\varepsilon/\Delta x$, the action $S=\int dxdt\left(
J+\partial_{x}\rho\right)  ^{2}/\left(  2\sigma\right)  $ is, to order
$O\left(  \varepsilon^{-1}\right)  $,%
\[
S=\frac{\Delta x\Delta t}{2\sigma_{0}\varepsilon^{2}}\left(  \frac{m}{\Delta
t}+\frac{\varepsilon}{\Delta x}\right)  ^{2}=\frac{1}{2\sigma_{0}%
\varepsilon^{2}}\left(  m^{2}v+\frac{\varepsilon^{2}}{v}+2m\varepsilon\right)
\]
where $v\equiv\frac{\Delta x}{\Delta t}$. We distinguish between two cases:
$J$ \textquotedblleft uphill\textquotedblright (against Fick's law,
requiring a strong noise), i.e. $sign\left(  J\right)  =sign\left(
\partial_{x}\rho\right)  $, and \textquotedblleft downhill\textquotedblright%
\ with $sign\left(  J\right)  =-sign\left(  \partial_{x}\rho\right)  $.
In\ the \textquotedblleft uphill\textquotedblright\ case, $\frac{m}{\Delta t}$
and $\frac{\varepsilon}{\Delta x}$ have the same sign, and minimizing $S$ over
$v\,$\ we obtain $v=\varepsilon/m$ or $S\geq\frac{2m}{\sigma_{0}\varepsilon
}+O\left(  \varepsilon^{0}\right)  $. To push a macroscopic mass $M$, the
bound will then read $S\geq\frac{2M}{\sigma_{0}\varepsilon}+O\left(
\varepsilon^{0}\right)  $. In contrast, if $J$ is \textquotedblleft
downhill\textquotedblright, $J+\partial_{x}\rho$ can be made small, with no
bound of order $O\left(  \varepsilon^{-1}\right)  $.

We now argue that for some profiles $\rho_{f}$, such as the one depicted in
Fig. \ref{fig:sweater_sleeve}(c,d), the global minimizing history is not
unique. We consider boundary conditions $\rho_{L}<-\varepsilon/2$ and
$\varepsilon/2<\rho_{R}$, so that $\rho_{f}$ crosses the noise barrier three
times. To highlight the symmetry-breaking aspect of the phenomenon, we focus
on a model with a $%
\mathbb{Z}
_{2}$ symmetry where $\sigma_{1}\left(  \rho\right)  =\sigma_{1}\left(
-\rho\right)  $, $\rho_{L}=-\rho_{R}$, and on a final profile satisfying
$\rho_{f}\left(  x\right)  =-\rho_{f}\left(  1-x\right)  $. Then
$\sigma\left(  \rho\right)  ,D$ and $\rho_{f}$ have a symmetry under the
combined operation of $\rho\rightarrow-\rho$ and $x\rightarrow1-x$. Let
$\rho^{sym}\left(  x,t\right)  $ be the minimizer of $S$ subject to this
symmetry. In the space of histories it is extremal, but not necessarily the
minimizing history. Indeed, we show that other solutions have lower action,
spontaneously breaking the symmetry. Referring to Fig.
\ref{fig:sweater_sleeve}(c) we note that in a symmetric history $\rho\left(
x,t\right)  $ we have $\rho\left(  x=1/2,t\right)  =0$, and masses in the
region shaded in gray, $2M$, must by pushed \textquotedblleft
uphill\textquotedblright\ through the $\rho=0$ line, hence $S\geq\frac
{4M}{\sigma_{0}\varepsilon}$. However, in the symmetry-breaking history shown
in Fig. \ref{fig:sweater_sleeve}(d) only a mass $M$ has to be pushed
\textquotedblleft uphill\textquotedblright, so that $S=\frac{2M}{\sigma
_{0}\varepsilon}+O\left(  \varepsilon^{0}\right)  $. Therefore, for a small
enough $\varepsilon$ this solution is favorable. By symmetry, the solution
with $\rho\rightarrow-\rho$ and $x\rightarrow1-x$ has the same action.
Therefore the symmetry is spontaneously broken, leading to the LDS with a
switching line on profiles obeying this symmetry. Note that some profiles
$\rho_{f}$, such as those in Fig. \ref{fig:sweater_sleeve}(b), can be
generated by a history which only pushes mass \textquotedblleft
downhill\textquotedblright\ through $\rho\in\left[  -\varepsilon
/2,\varepsilon/2\right]  $, and the above argument for multiple histories will
not hold. While the above argument emphasizes the breaking of a symmetry, the
phenomenon exists even in the absence of an explicit symmetry in either the
final configuration or the model, as was demonstrated in the first two models considered.

The non-differentiability shown in Figs. \ref{fig:exact_model}(c) and
\ref{fig:KLS}(c,d) is perhaps the simplest possible singularity
\cite{Dykman_cusp_structure}. Due to the high dimensionality of the phase
space of profiles, more elaborate structures are possible. Understanding and
classifying the possible singularities in these models is an exciting direction
for future research.

\emph{Acknowledgments} - This research was funded by the BSF, ISF, and IRG grants.


\begin{thebibliography}{99}
\bibitem {all_quad_sig}It is easy to show that the phenomenon exists in all
models with $D=1$ and $\sigma\left(  \rho\right)  =a+b\rho^{2}$ with $a,b$ positive.
\end{thebibliography}
\end{document}